\documentclass[preprint]{aastex}

\newcommand{\etal}{et~al.}
\newcommand{\kms}{km~s$^{-1}$}

\newcommand{\CIVdblt}{\ion{C}{4}~$\lambda\lambda 1548, 1550$}
\newcommand{\CII}{\ion{C}{2}}
\newcommand{\CIV}{\ion{C}{4}}
\newcommand{\SiIV}{\ion{Si}{4}}
\newcommand{\NV}{\ion{N}{5}}
\newcommand{\OVI}{\ion{O}{6}}
\newcommand{\HI}{\ion{H}{1}}
\newcommand{\Lya}{{Ly}\,$\alpha$}
\tighten
\begin{document}

\slugcomment{{\it The Astrophysical Journal Letters}}
\shortauthors{~~Ganguly \etal}
\shorttitle{CSO 118 Absorbers}

\title{
\bf The Absorbers Toward CSO 118: Superclustering at {z$\sim3$}, or an
Intrinsic Absorption Complex?\footnote{Based on observations obtained with the Hobby-Eberly Telescope, which is a joint project of the University of Texas at
Austin, the Pennsylvania State University, Stanford University,
Ludwig-Maximillians-Universit$\ddot{\mathrm{a}}$t
M$\ddot{\mathrm{u}}$nchen, and
Georg-August-Universit$\ddot{\mathrm{a}}$t
G$\ddot{\mathrm{o}}$ttingen.}}

\pagestyle{empty}

\author{Rajib~Ganguly, Jane~C.~Charlton, Nicholas~A.~Bond}
\medskip
\affil{\normalsize\rm Department of Astronomy and Astrophysics \\
       The Pennsylvania State University, University Park, PA 16802 \\
       e-mail: {\tt ganguly, charlton, bond@astro.psu.edu}}
\pagestyle{empty}
\begin{abstract}
We present two low resolution ($R\sim1300$) high signal-to-noise
spectra ($W_{\mathrm{r,lim}}\approx50$~m\AA) of the quasar CSO 118
($z_{\mathrm{em}}=2.97$) taken with the Hobby--Eberly Telescope
Marcario Low Resolution Spectrograph. We detect eight absorbers
selected by the {\CIVdblt} absorption doublet in the redshift range
{$2.23 \lesssim z_{\mathrm{abs}} \lesssim 2.97$}, seven of which are
{$z_{\mathrm{abs}} \gtrsim 2.68$}. In the redshift range covered by
the seven {$z_{\mathrm{abs}} \gtrsim 2.68$} systems, one expects to
find two absorbers. We discuss possible explanations for such an
excess of absorbers in a small velocity range. Only superclustering at
high redshift and absorption due to intrinsic gas are feasibly
allowed.
\end{abstract}

\keywords{quasars: absorption lines -- quasars: individual (CSO 118)}

\section{Introduction}
\pagestyle{myheadings}

In recent years, the analysis of QSO-intrinsic narrow absorption lines
(NALs) has become a productive enterprise. There are two smoking guns
for the identification of intrinsic NALs: (1) time variability of
profile shapes and/or equivalent widths; and (2) demonstration that
the absorbing structures only partly occult the background source (the
QSO central engine). The latter option can only viably be pursued
using high resolution and reasonably high signal-to-noise spectra
\citep{bs97,bhs97,ham97a,gan99,sp00}. In the former case, we can
search for variability in the equivalent width of absorption profiles
using low resolution spectra. To that end, we have embarked on a
monitering program of the Ver\'on-Cetty \& Ver\'on QSOs using the
Marcario Low Resolution Spectrograph on the Hobby--Eberly
Telescope. Finding variability in absorption profile equivalent widths
is a step toward systematically identifying truly intrinsic
NALs. Follow-up observations at higher resolving power to look for
changes in profiles shapes and/or the signature of partial coverage
will remove any shadow of a doubt as to an intrinsic origin.

In this letter, we report on a curiosity - the serendipitous discovery
of a complex of absorbers that are possibly intrinsic to the
radio--quiet quasar CSO 118 ({$z_{\mathrm{em}}=2.97$}, {$V=17.0$}). In
\S2, we present the spectra of CSO 118 and details of the
observations. In \S3, we demonstrate that the absorption complex is
unlikely to be a random occurrence. Finally, in \S4 and \S5, we discuss
the possible explanations for this complex under the assumptions that
the complex is intervening or intrinsic, respectively.

\section{Observations}
\begin{figure}
\figurenum{1}
\plotone{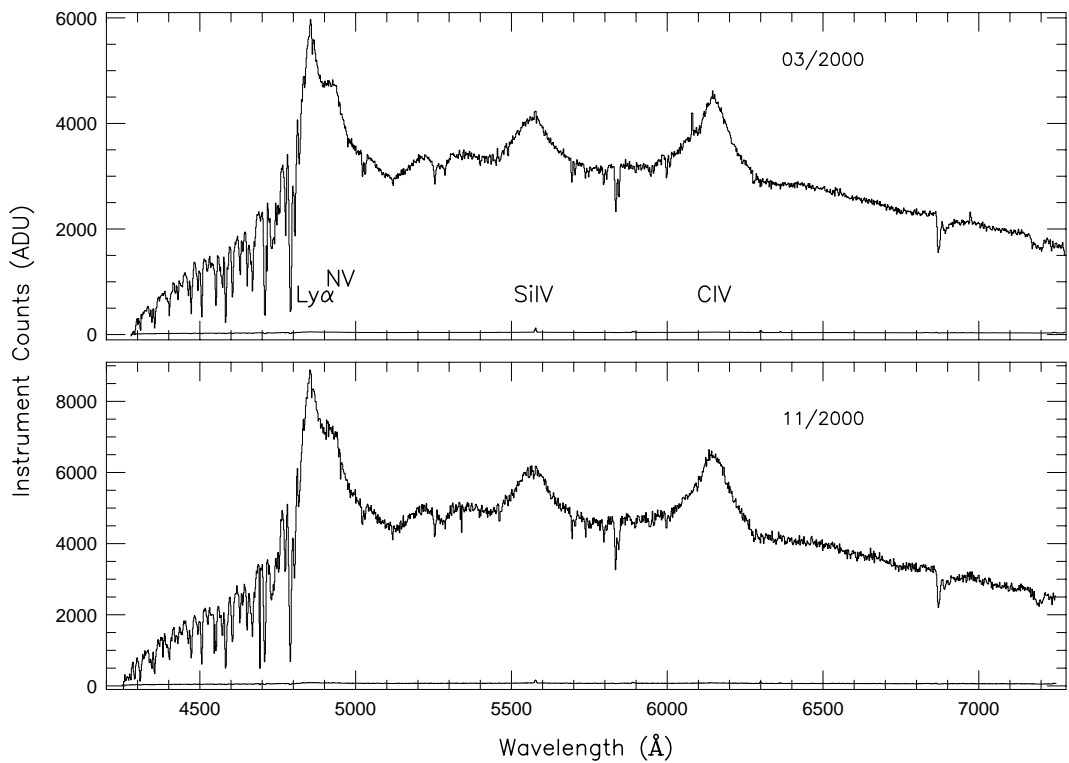}
\vglue -0.5in
\figcaption[fig1.eps]{
Spectra of CSO 118 taken at two epochs (March and November 2000) with
the Marcario Low Resolution Spectrograph \citep{lrs}. The spectra
cover the {Ly$\alpha$}, N {\sc v}, Si {\sc iv}, and C {\sc iv}
emission lines, and part of the {Ly$\alpha$} forest.  The spectra have not
been flux calibrated or corrected for atmospheric absorption, since
HET is set at a constant airmass ($\sec z = 1.1-1.3$). We detect one C
{\sc iv}--selected absorption system on the red wing of the N {\sc v}
emission and seven between the Si {\sc iv} and C {\sc iv} emission
lines.}
\label{fig:bigspec}
\end{figure}

\begin{figure}
\figurenum{2}
\plotone{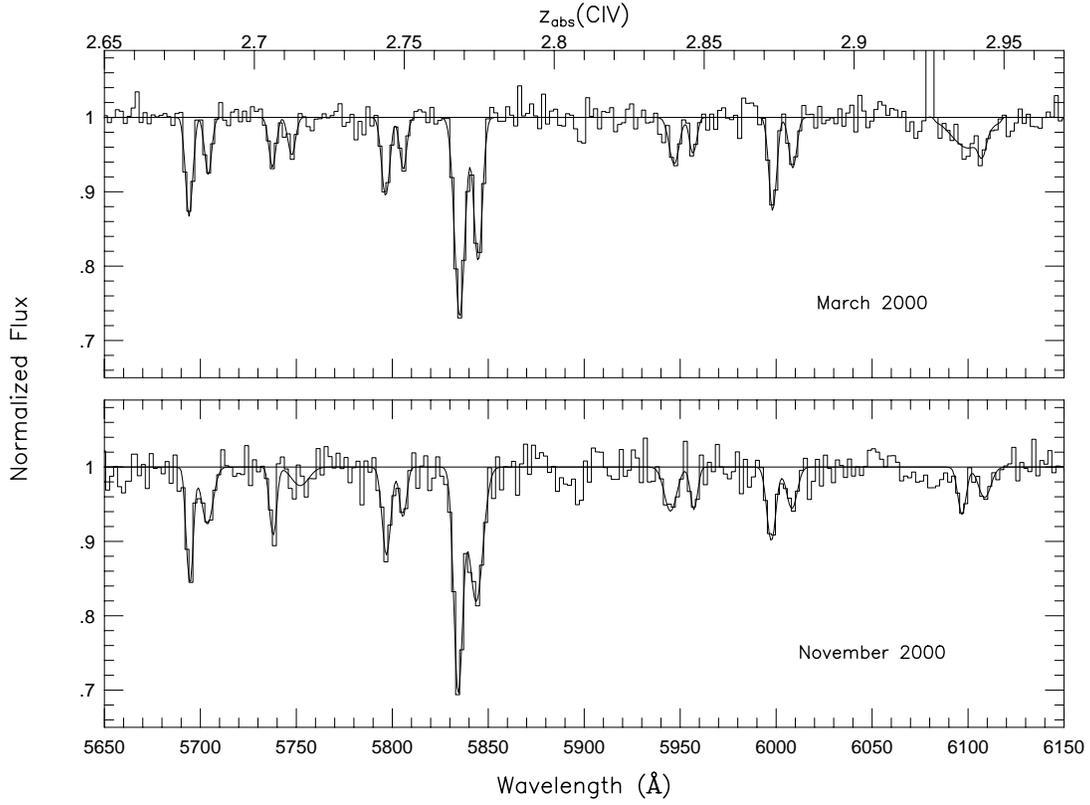}
\vglue -0.5in
\figcaption[fig2.eps]{
A ``zoomed'' in region, from 5650--6150 \AA, of the spectra of CSO
118. Overlayed on the data are the double Gaussian fits to the each of
the C {\sc iv} doublet profiles. On the top axis, we show the redshift
of the C {\sc iv} {$\lambda1548$} line. This part of the spectrum
shows the seven C {\sc iv}--selected absorbers in the redshift range
{$z_{\mathrm{abs}}=2.68-2.94$}. The equivalent widths and doublet
ratios of the profiles did not change during this rest--frame two
month period.}
\label{fig:zoomspec}
\end{figure}

Two spectra were obtained, separated by eight months, with the
Hobby-Eberly Telescope (HET) using the Marcario Low Resolution
Spectrograph \citep{lrs}.  We used the 600 line/mm grism and the 1''
slit to achieve a resolving power of {$R\sim1300$} or {$\Delta
v\approx230~\mathrm{km~s}^{-1}$} resolution. In
Table~\ref{tab:journal}, we report the observing dates and the
signal-to-noise (per pixel) of the spectra. The spectra were bias
subtracted, and flat-fielded using the standard IRAF\footnote{IRAF is
distributed by the National Optical Astronomy Observatories, which are
operated by the Association of Universities for Research in Astronomy,
Inc., under cooperative agreement with the National Science
Foundation.} image reduction packages. Spectra were extracted using
the {\sc apall} task and wavelength calibrated. Due to the varying
aperture size of the HET, we opted not to attempt flux
calibration. Also, since the HET is set at a constant zenith angle,
the two spectra suffer from the similar amounts of atmospheric
absorption. Thus, it was not necessary to correct for differing
airmasses. The spectra cover the range {$4280-7270$~\AA}. In each
spectrum, the {$\sim0.5$~hr} integration time gives a {$3\sigma$}
rest--frame equivalent width limit better than {$\sim50$~m\AA}, which
we also report in Table~\ref{tab:journal}. In Fig.~\ref{fig:bigspec},
we show the two spectra as instrument counts versus wavelength. The
emission lines from {\Lya}, {\NV}, {\SiIV}, and {\CIV} are clearly
visible as well as the {$2.52<z_{\mathrm{abs}}<2.97$} {\Lya} forest
absorption blueward of the {\Lya} emission line. Using the unresolved
feature identification method of \citet{kpii} and \citet{cwc99}, we
identified possible {\CIV} doublets where both the stronger transition
($\lambda1548$) and the corroborating {\Lya} were detected at a
{$3\sigma$} confidence and the weaker {\CIV$\lambda1550$} doublet
transition at {$1.5\sigma$} confidence [see e.g. \citet{gan01}]. We
detect eight {\CIV}--selected absorbers in the redshift range
2.23--2.97, seven of which are within {$\Delta z=0.26$} of the QSO
emission redshift. To each doublet, we fit two Gaussians to measure a
deblended {\CIV$\lambda1548$} equivalent width. We list these in
Table~\ref{tab:ew} for all eight systems over both epochs of
observation as well as the doublet ratios
($=W_{\mathrm{r}}(1548)/W_{\mathrm{r}}(1550)$). In only one case
({$z_{\mathrm{abs}}=2.94$} on {$4~\mathrm{March}~2000$}) was
deblending unsuccessful due to an insufficiently resolved doublet. In
Fig.~\ref{fig:zoomspec}, we show the region of the spectra covering
the wavelength range 5650--6150 {\AA}. The spectra have been
normalized so that the flux from both the continuum and emission lines
(i.e., the total emitted flux indicent on the absorbers) is
unity. Overplotted, we show the fits of the double Gaussians to the
seven {\CIV} profiles in this wavelength range.

\section{Redshift Path Density of Absorbers}

{\citet{tls96}} reported that the redshift path density of {\CIV}
absorbers in the redshift range {$1.4<z_{\mathrm{abs}}<2.9$} and down
to an equivalent width limit of {30m\AA} is {$dN/dz=7.1\pm1.7$}.  The
total redshift path searched in the spectrum of CSO 118 is {$\Delta
z=0.85$}, so we would expect to find {$6.0\pm1.4$} in all. The Poisson
probability of finding eight systems when six are expected is 10\%.
However, seven of the absorbers are clustered on the high redshift
side of this path. If one divides the path in two, one finds one
absorber in the low reshift bin, and seven in the high redshift
bin. In each of these bins, the redshift path is {$\Delta z=0.425$} in
which one expects to find {$3.0\pm0.7$} absorbers on average. The
Poisson probabilities of finding one and seven absorbers in the
{$\Delta z=0.425$} path are {$14\pm7\%$} and {$2\pm2\%$},
respectively. Finding only one absorber in the low redshift bin is not
a statistically significant decrement. However, finding seven in the
high redshift bin is a very significant excess. Moreover, if such an
absorption complex were common, one would expect the two-point
correlation function (TPCF) of {\CIV}--selected systems at {$z\sim3$}
to show an above average amplitude at large
velocities. \citet{rauch96} report that the amplitude of the TPCF
falls off to the average value (i.e., uncorrelated distribution of
systems in redshift space) beyond {$400$~\kms}. Therefore, this
complex is not only highly significant, but also very rare.  We
discuss two possible origins for the excess of {\CIV} doublets in the
spectrum of CSO 118: (1) absorption from intervening structures; or
(2) absorption by gas intrinsic to the QSO.

\section{Intervening Absorption}

Intervening {\CIV} absorption has traditionally been attributed to
galaxy halos \citep{ssb87,pb94}, high ionization species associated
with the {\Lya} forest \citep{luthesis,sc96,kt99}, or hierarchical
galaxy formation [\citet{rhs97}; hereafter RHS]. The hierarchical
collapse of {\HI} into sheets and filaments has been shown through
many impressive simulations [e.g., \citet{dave99,mac00}] to reproduce
the so-called {\Lya}--forest which is prevalent in the spectra of QSOs
blueward of the {\Lya} emission.

Both galaxy halo absorption and forest absorption (from {\CIV}) are
accounted for by standard {$dN/dz$} measurements. Thus only the
presence of a few absorbers ($3.0\pm0.7$~in~$\Delta z=0.425$) can be
explained in this manner. In the model of hierarchical galaxy
formation, as discussed by RHS, a dense {\HI} filament (see, for
example, their Fig. 3) gives rise to low ionization absorption (e.g.,
{\HI}, {\CII}, {\SiIV}).  This is surrounded by a lower density phase
(giving a high ionization parameter) that produces {\CIV}
absorption. This lower density phase is encompassed by an even lower
density phase that yields very high ionization {\OVI} absorption. A
HIRES/Keck spectrum of the {$z_{\mathrm{abs}}=2.768$} system
[\citet{kt99}; hereafter KT] toward CSO 118 shows precisely this type
of structure (see their Fig. 1). KT performed a Voigt profile
decomposition of the system profiles to extract column densities and
Doppler widths for each velocity component of each transition. Noting
the differences in the {\CIV} and {\OVI} Doppler widths, KT reported
this as evidence of distinct high ionization phases and speculated
that this was evidence of a hierarchical merging event.

While the hierarchical galaxy formation scenario may also be
responsible for a few absorbers, it is unlikely to explain a whole
complex spread over {$20,000$~\kms} as seen along the CSO 118 line of
sight. Regardless of whether or not the absorption is intrinsic or
intervening, the seven {$2.68<z_{\mathrm{abs}}<2.94$} absorption
systems must arise from structures that are somehow
correlated. Another possibility related to the RHS scenario is the
fragmentation of a large {\HI} filament to from several
``protogalactic clumps'' (PGCs), that is, the formation of a cluster of
galaxies or a supercluster.  The seven absorbing systems are spread
over {$20,000$~\kms}. Moreover, rich clusters of galaxies typically
have velocity dispersions of {$\sim1000$~\kms}; it is rare, indeed, to
find clusters with dispersions as large as {$2000$~\kms} -- an order
of magnitude smaller than what is required here. The only viable
explanation invoking intervening absorption is superclustering.

Superclustering involves the clustering the galaxy clusters. Several
superclusters have been identified in the last couple decades like the
Great Wall \citep{cfa1,cfa2} from the CfA redshift surveys (which
includes the Coma cluster), the Perseus--Pisces supercluster
\citep{gtt81}, the Hercules supercluster \citep{tar79,crt81,gt84},
and the Local Supercluster \citep{yst80,tully82,huchra83}. The bulk of
these low redshift superclusters are dominated by rich clusters like
those in the \citet{abell58} and \citet{aco89} catalogs. In instances
where the filamentary structure seen in these superclusters is aligned
with the line of sight such as in the Aquarius supercluster
\citep{bat99}, the velocity distribution of clusters can reach
{$1-2\times10^4$~\kms}. At high redshift, if the corresponding
``protocluster'' clumps (PCCs) provide {\CIV} absorption in the
spectrum of a background QSO, this could explain the complex of
absorption seen in the CSO 118 line of sight. It is not difficult to
imagine an {\HI} filament that is fragmenting into PCCs with a QSO
behind it. The extragalactic background, dominated at high redshift by
the ultraviolet and reprocessed soft X--ray emission from AGN
\citep{hm96}, ionizes the PCCs. The ionization structure of each of
these PCCs would be similar to that described by RHS and KT, with
multiple phases of low density, high ionization gas surrounding higher
density, lower ionization gas.

Superclustering is not a new idea in the realm of quasar absorption
lines.  There have been several reports in the literature of binary
QSOs in which absorption at common redshifts occur. The most famous of
these is the ``Tololo pair,'' {Tol~$1037-2704$} and {Tol~$1038-2712$}
\citep{bw79,jak86}, where the candidate supercluster is at {$z\sim2$}.
Other reports include UM~680/681, and Q~$2343+125$/Q~$2344+125$
{\citep{sar88}} [see also \citet{rfs91,fh93,wil96,fra96,wil00}]. These
claims are usually based on the coincidence of absorbers in redshift
(implying gas that is possibly in common to both lines of sight) and
the separation of the QSOs (17' in the case of the Tololo pair
implying a {\it minimum} size of
{$\sim4\mathrm{h}^{-1}$~Mpc}). However, these claims are still
controversial as it is still unknown whether they caused by
intervening structures or whether they arise due to the presence of a
relativistic outflow from the QSO central engine (see \S5). We note
here that there are no bright extragalactic objects (for which to do
absorption line spectroscopy) in the NASA/IPAC Extragalactic Database
at high redshift within 20' of CSO 118. Thus, a study of the
transverse extent of this structure (if it is intervening) is
unlikely. There is, at least, one other object in the literature,
{Q~$2359+068$}, with a similar concentration of absorbers (8 in the
range {$2.73 < z_{\mathrm{abs}} < z_{\mathrm{em}}$}, with a Poisson
probability of {$1\pm1$\%}).

\section{Intrinsic Absorption}

The complex of high redshift absorbers toward CSO 118 can also
plausibly be of an intrinsic origin. First, intrinsic NALs have been
detected at high ejection velocities up to
{$v_{\mathrm{ej}}\sim56000$~km/s} \citep{jan96,ham97b,gtr99}.  Second,
multiple (2--3) intrinsic absorption systems are also observed --
e.g., {PG~$2302+029$} \citep{jan96}, {PG~0935+417}
\citep{ham97c}, {Q~$0835+5803$} \citep{ald97}. Furthermore, at low 
redshift, there is an enhanced probability of a radio--quiet QSO
hosting an intrinsic NAL when one detects broad absorption (a BAL)
\citep{gan01}. Thus, it is reasonable to expect multiple
absorption systems that are intrinsic to a given QSO.

Unfortunately, on the two month rest--frame timescale over which the
CSO 118 was observed, neither the equivalent width, nor the doublet
ratio of any profile varied. (Also, since the features are unresolved,
it make little sense to compute coverage fractions.) Either evidence
for time variability or partial coverage would provide a smoking gun
for an intrinsic origin. Nevertheless, it must be noted that the {\it
lack} of variability of the profiles does not preclude an intrinsic
origin. Moreover, because the profiles are unresolved, we would not
have detected changes in the profile shapes. We offer two possible
scenarios to explain this absorption complex under the
accretion-disk/wind model \citep{mur95,psk00}. The first scenario
involves the presence or the development of a BAL outflow, while the
second proposes periodic or stochastic mass loss events.

In the accretion-disk/wind model for the QSO central engine, matter
orbits a supermassive black hole in a geometrically thin, optically
thick disk and spirals inward as a result of viscous friction. Matter
on the surface of the disk is lifted via radiation pressure. The rate
at which this matter is lost is regulated by the balance between the
mass accretion rate, which is capped by the Eddington rate, and the
mass fueling rate. As this mass is blown off the disk, it is
blindsided by an even more powerful force: radiation pressure from the
inner part of the disk which radiates a UV/soft X--ray continuum. This
results in a relativistically and radially accelerated wind blowing
away from the disk. However, the matter leaving the disk retains its
angular momentum. So as it is radially accelerated, it spirals away
from the disk in a helix. This rotational component can play an
important part in the projected line-of-sight velocity of the wind
and, by consequence, the optical thickness to photons emitted by the
accretion disk and broad line region. The broad UV emission lines
originate from the lower regions of the wind where the matter is
optically thick.

The complex of absorption seen toward CSO 118 may be connected to
BAL-type outflow. In the accretion-disk/wind model described above,
BAL outflow is understood to occur when the mass fueling rate greatly
exceeds the mass accretion rate. In this case, the wind becomes
optically thick and photons attempting to pass through the wind are
completely absorbed. Moreover, since there is a large velocity
gradient in the wind, the absorption, as well, occurs over a broad
range in velocities. Thus, all radio--quiet QSOs are viewed as {\it
having} a BAL outflow, but in only {$\sim10$\%} of such cases are the
viewing angle and wind opening angle such that a BAL is seen. CSO 118
fits in nicely with being like a BALQSO in which the line(s) of sight
graze the BAL outflow. The absorbers span ejection velocities up to
{$23,000$~\kms}, reminiscent of typical BAL velocity widths. Like many
BALQSOs, CSO 118 is very radio--quiet (a non-detection by the FIRST
survey makes the radio--loudness parameter $\log R<0$). It is also not
detected by the ROSAT All Sky Survey (although this merely provides
the unrestrictive limit of {$\alpha_{\mathrm{ox}}\lesssim-1.3$}). One
possible explanation for the difference between the series of NALs in
CSO 118 and a BAL is that the line of sight to the latter grazes
clumpy outcroppings of the wind; these clumps, which are cause by
Kelvin--Helmholtz instabilities, are seen in the simulations of Proga
{\etal}~(2000). According to the simulations, these clumps last on the
order of a few years in the QSO rest--frame. A related possibility is
that the outflowing wind is starting to become much denser, possibly
as a result of a changeins mass fueling/loss rate. It is possible
that we are seeing the initial fragmentation of the wind into BAL
clouds.

Another possible explanation for CSO 118 is sporadic (or
quasi--periodic) mass ejection by the accretion disk.  Because the
mass accretion rate is capped by the Eddington limit, changes in the
mass fueling rate are directly transposed to changes in the mass loss
rate. If the fueling rate were to drastically change either
periodically or sporadically, the outflowing wind would have a density
structure, which can be viewed as a perturbations on the general
density law of the wind. The perturbations resulting from an increase
in the mass fueling rate would resemble expanding shells. The
observational signature of outflowing shells could be complexes of
absorbers as seen in the CSO 118 line of sight.

\acknowledgements

Support for this work was provided by the NSF (AST-9617185) and NASA
(NAG5-6399). We would like to thank Gary Hill for building the LRS and
Larry Ramsey for the building the Hobby--Eberly Telescope. RG
acknowledges Gordon Richards for useful discussions.


\clearpage
\begin{deluxetable}{cccc}
\tablewidth{0pc}
\tablenum{1}
\tablecaption{Journal of Observations of CSO 118}
\tablehead
{
\colhead{Observation} &
\colhead{Exposure} &
\colhead{} &
\colhead{$3\sigma~W_{\mathrm{r}}$~limit} \\
\colhead{Date} &
\colhead{Time} &
\colhead{$S/N$} &
\colhead{m\AA}
}
\startdata
$03/04/2000$ & 1800s & 70 & 36 \\
$11/21/2000$ &  600s &    &    \\
$11/24/2000$ &  900s & 55\tablenotemark{a} & 45\tablenotemark{a} \\
$11/25/2000$ &  900s &    &    \\
\enddata
\tablenotetext{a}{The three integrations taken in November {$2000$}
were co-averaged to give the reported {$S/N$} and {$3\sigma~W_{\mathrm{r}}$}
limit. The smaller {$S/N$} in the November spectrum resulted primarily from
poorer seeing conditions.}
\label{tab:journal}
\end{deluxetable}

\begin{deluxetable}{ccccccc}
\tablewidth{0pc}
\tablenum{2}
\tablecaption{Equivalent Widths of {\CIV} doublets}
\tablehead
{
\colhead{} &
\multicolumn{3}{c}{$03/04/2000$} &
\multicolumn{3}{c}{$11/24/2000$} \\
\colhead{} &
\multicolumn{3}{c}{\hrulefill} &
\multicolumn{3}{c}{\hrulefill} \\
\colhead{$z_{\mathrm{abs}}$} &
\colhead{Wavelength} &
\colhead{$W(1548)$} &
\colhead{D.R.} &
\colhead{Wavelength} &
\colhead{$W(1548)$} &
\colhead{D.R.} \\
\colhead{} &
\colhead{\AA} &
\colhead{\AA} &
\colhead{} &
\colhead{\AA} &
\colhead{\AA} &
\colhead{}
}
\startdata
$2.244$ & $5022.4\pm0.2$ & $0.59\pm0.06$ & $0.9\pm0.1$ & $5021.6\pm0.3$ & $0.67\pm0.08$ & $1.5\pm0.3$ \\
$2.678$ & $5694.2\pm0.2$ & $0.63\pm0.06$ & $1.8\pm0.3$ & $5694.5\pm0.3$ & $0.71\pm0.07$ & $1.3\pm0.4$ \\
$2.705$ & $5737.6\pm0.4$ & $0.31\pm0.06$ & $1.3\pm0.4$ & $5737.9\pm0.5$ & $0.42\pm0.07$ & $1.5\pm1.0$ \\
$2.744$ & $5796.6\pm0.3$ & $0.56\pm0.07$ & $1.8\pm0.4$ & $5797.0\pm0.4$ & $0.67\pm0.08$ & $2.1\pm0.8$ \\
$2.769$ & $5835.1\pm0.1$ & $1.80\pm0.08$ & $1.7\pm0.1$ & $5834.3\pm0.2$ & $1.74\pm0.09$ & $1.1\pm0.2$ \\
$2.841$ & $5946.9\pm0.5$ & $0.41\pm0.08$ & $1.8\pm0.6$ & $5945.0\pm1.0$ & $0.48\pm0.10$ & $1.8\pm0.9$ \\
$2.874$ & $5998.2\pm0.2$ & $0.63\pm0.06$ & $1.8\pm0.4$ & $5997.5\pm0.5$ & $0.58\pm0.08$ & $1.5\pm0.6$ \\
$2.940$ & $6100.1\pm1.9$ & $0.89\pm0.14$\tablenotemark{a}
                                         & \nodata     & $6096.8\pm0.7$ & $0.34\pm0.07$ & $1.0\pm0.5$ \\
\enddata
\tablenotetext{a}{The {\CIV} doublet profile could not be deblended. The
listed equivalent width is of the blended feature.}
\label{tab:ew}
\end{deluxetable}

\clearpage

\end{document}